\documentclass[10pt,letterpaper,twocolumn]{article} 
\usepackage{ol2}

\usepackage[draft]{hyperref}
\usepackage{amsmath}

\begin{document}

\twocolumn[ 

\title{Continuous-wave spontaneous lasing in mercury pumped by resonant two-photon absorption}


\author{Daniel Kolbe,$^*$ Martin Scheid, Andreas Koglbauer, and Jochen Walz}

\address{
Institut f{\"{u}}r Physik, Johannes Gutenberg-Universit{\"{a}}t Mainz and Helmholtz-Institut Mainz,  D-55099 Mainz, Germany

$^*$Corresponding author: kolbed@uni-mainz.de
}

\begin{abstract}
The first continuous-wave two-photon absorption laser-induced stimulated emission (CTALISE) is demonstrated. The 7$^1S_0$--6$^1P_1$ transition in mercury at 1014\,nm wavelength is used and selective lasing of different isotopes is observed.
\end{abstract}


] 

\noindent 

Multiphoton processes are at the heart of many applications of photonics and quantum optics. Fascinating examples include upconversion lasers \cite{Hebert90}, multiphoton microscopy \cite{Bewersdorf98} and data storage in polymers \cite{Strickler91}. Two-photon absorption can even drive spontaneous lasing, a process called TALISE (two-photon absorption laser-induced stimulated emission) \cite{Goldsmith89}, which can be used for plasma diagnostics \cite{Amorim00}. Such two-photon pumped amplified spontaneous emission (ASE) was observed in mercury on the $7^3S_0$--$6^3P_1$ transition at 546\,nm \cite{Omenetto94}. The cross-section of multiphoton processes is often small compared to one photon processes. High intensities are therefore necessary to reach enough gain for lasing which could only be achieved with pulsed lasers, so far.

This Letter reports the first two-photon pumped continuous-wave laser. The two-photon excitation is enhanced by a near one-photon resonance. Continuous-wave laser powers of about 10--100\,mW are therefore sufficient to induce amplified spontaneous emission. We investigate threshold conditions for different mercury isotopes and compare two different pumping schemes as a function of buffer gas pressure in the mercury cell.

\begin{figure}[tb]
\centerline{\includegraphics[width=4.2cm]{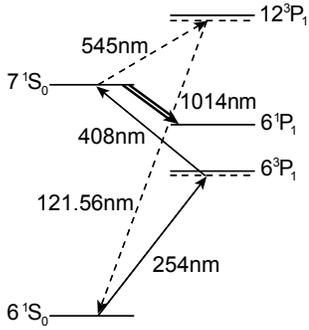}}
\caption[Level scheme]{Energy-level diagram of mercury. The UV laser (254\,nm) is tuned close to the $6^1S_0$--$6^3P_1$ resonance, the blue laser (408\,nm) establishes the two-photon resonance with the $7^1S_0$ state. Lasing is observed at the $7^1S_0$--$6^1P_1$ transition (1014\,nm). The green laser (545\,nm) can be used in addition to produce radiation at Lyman-$\alpha$ (121.56\,nm) by four-wave-mixing.}
\label{Fig:1}
\end{figure}

Figure \ref{Fig:1} shows a simplified level scheme of mercury and the pumping transitions at 254\,nm and 408\,nm wavelength. Lasing occurs on the $7^1S_0$--$6^1P_1$ transition at 1014\,nm wavelength. In this scheme an additional green laser at 545\,nm can be used to produce Lyman-$\alpha$ radiation (121.56\,nm) by four-wave-mixing \cite{Scheid09}. Radiation at 254\,nm comes from a frequency quadrupled Yb:YAG disc-laser which generates up to 750\,mW in the UV \cite{Scheid07}. Radiation at 408\,nm comes from a frequency doubled titanium:sapphire laser with an output power of up to 450\,mW in the blue. The pumping beams are focused by a lens of 15\,cm focal length into a mercury cell with a vapor region of about 1.5\,cm length. The infrared light is separated from the pumping beams by an interference filter at 1014\,nm and detected with a photodiode. Details of the mercury cell and the detection method are given elsewhere \cite{Eikema01}. The beam waists in the focus are 8\,$\mu$m (UV) and 10\,$\mu$m (blue).

The detuning of the two lasers and the intensities of the laser fields can be used to control the population in the upper $7^1S_0$ level. Its lifetime of 32\,ns is long compared to the vacuum lifetime of the lower $6^1P_1$ level (1.5\,ns) \cite{Benck89}, although latter may be significantly increased by radiation trapping in dense mercury vapor \cite{Menningen00}. Pumping into the upper level thus produces a population inversion on the $7^1S_0$--$6^1P_1$ transition, which yields amplified spontaneous emission. To enhance the two-photon excitation we choose a detuning in the UV of \mbox{-25\,GHz} relative to the $6^1S_0$--$6^3P_1$ one-photon resonance of the $^{202}$Hg isotope. At smaller detunings resonant absorption of the UV beam is too large to efficiently pump the mercury vapor along the entire length of the cell.

\begin{figure}[tb]
\centerline{\includegraphics[width=5.8cm,angle=-90]{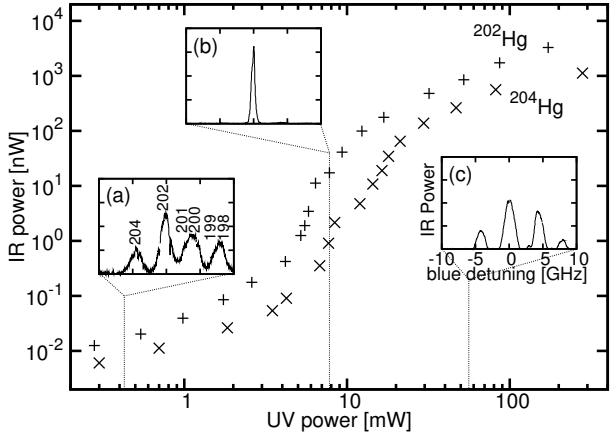}}
\caption[Threshold]{Threshold behavior of CTALISE. The infrared power emitted in foreward direction is shown as a fuction of UV pump power at the two-photon resonance of $^{202}$Hg (+) and of the $^{204}$Hg isotope (X). The insets show scans with the same span across the two-photon resonance in the three regimes: (a) fluorescence, (b) threshold and (c) lasing.}
\label{Fig:2}
\end{figure}

Figure \ref{Fig:2} shows the power of the infrared light as a function of the UV power at the two-photon resonance of the $^{202}$Hg and $^{204}$Hg. The temperature of the mercury cell is 120\,$^\circ$C (mercury density $N_{\textrm{Hg}}=1.8\times 10^{22}$m$^{-3}$), the power of the blue beam is 45\,mW and no buffer gas is used. The insets show scans of the blue frequency across this resonance at different UV powers.  Three different regimes can be distinguished and will be discussed in more detail in the next section: the first at low UV powers up to 4\,mW (fluorescence), the second between 4\,mW and 20\,mW (threshold) and the third above 20\,mW (lasing).

In the fluorescence regime at low UV powers the IR signal is proportional to the UV power. The scan across the two-photon resonance (see inset (a) of Figure \ref{Fig:2}) shows several Voigt-profils resulting from the isotope splitting of the $6^1S_0$--$7^1S_0$ resonance.  The height of the peaks is related to the natural abundance of the isotopes ($^{204}$Hg: 6.9\%, $^{202}$Hg: 29.9\%, $^{201}$Hg: 13.2\%, $^{200}$Hg: 23.1\%, $^{199}$Hg: 16.9\%, $^{198}$Hg: 9.97\%) \cite{Zadnik89} and the linewidth of 1.5\,GHz is due to thermal Doppler broadening. The shape of the spectrum can be calculated using the optical Bloch equations of a three level system \cite{Beyer09}. 

In the threshold regime at intermediate UV powers the gain of the pumped vapor region is large enough so that lasing condition due to amplified spontaneous emission occurs. The $^{202}$Hg isotope has the highest abundance and thus the lowest threshold condition. At the threshold regime a steep increase in the IR power can be seen. Above the threshold of the $^{202}$Hg isotope but below the threshold of every other isotope the IR light is dominated by the contribution of the $^{202}$Hg isotope. The spontaneous fluorescence of the other isotopes is many orders of magnitude smaller. The scan shows a single narrow peak at the two-photon resonance of the $^{202}$Hg (Figure \ref{Fig:2}\,(b)). 

In the lasing regime at high UV powers all isotopes are above threshold and every isotope performs CTALISE as can be seen in the inset (c) of Figure 2. 
Comparing insets (a) and (b) of Figure 2 one clearly sees that the linewidth of the infrared signal is reduced in the regime of threshold. The broad excitation spectrum has to be multiplied with the nonlinear gain and is thereby narrowed.

\begin{figure}[t!]
\centerline{\includegraphics[width=5.8cm,angle=-90]{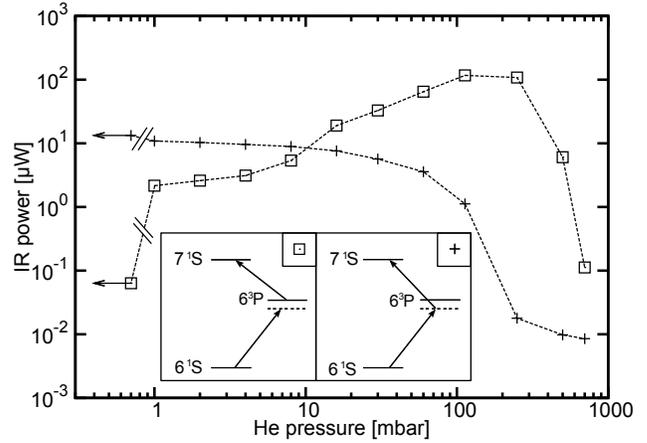}}
\caption[Pressure]{Infrared powers  at different buffer gas pressures. Two different pumping schemes are compared: two-photon resonant to the $^{202}$Hg (crosses) and double one-photon excitation of the $^{202}$Hg (boxes). Points with arrows were recorded at the lowest buffer gas pressure possible ($<10^{-3}$\,mbar).} 
\label{Fig:3}
\end{figure}

There are two different excitation schemes to the $7^1S_0$ level. The first one is a two-photon excitation. This means the sum of both laser frequencies is equal to the transition frequency of the $7^1S_0$--$6^1S_0$ transition. The second excitation scheme is a double one-photon excitation. In this case an offresonant exitation to the $6^3P_1$ level is followed by a resonant excitation into the upper $7^1S_0$ level. Therefore the blue laser frequency is at the frequency of the $6^3P_1$--$7^1S_0$ transition. In Figure \ref{Fig:3} we compare both pumping schemes at different buffer gas pressures in the mercury cell. The detuning of the UV laser is -25\,GHz. The pumping powers are about 200\,mW UV and 245\,mW blue and the cell temperature is 120\,$^\circ$C. Helium is used as buffer gas in a pressure range from 0 to 700\,mbar. The maximum laser power at 1014\,nm was 115\,$\mu$W at a buffer gas pressure of 100 to 250\,mbar and pumping with double one-photon excitation. Increasing buffer gas pressure causes more pressure broadening and thus to enhanced one-photon absorption at the UV wavelength. This effect has different influence on both excitation schemes: For the double one-photon excitation an enhanced UV absorption results in an enhanced excitation efficiency into the upper laser level. At buffer gas pressures above 300\,mbar the absorption at the UV wavelength is too high to efficiently pump the mercury along the entire cell and the infrared power decreases. For the two-photon resonant excitation, in contrast, there is no pressure related enhancement effect. UV absorption is a competing process instead. Increasing the buffer gas pressure decreases the UV power due to absorption. In addition, the density of Hg atoms in the ground state decreases due to excitation to the intermediate $6^3P_1$ level. The excitation probability by the competing two-photon process to the upper laser level decreases. Thus, the infrared power produced by the two-photon resonant excitation decreases as the helium pressures increases. 

\begin{figure}[tb]
\centerline{\includegraphics[width=8.3cm]{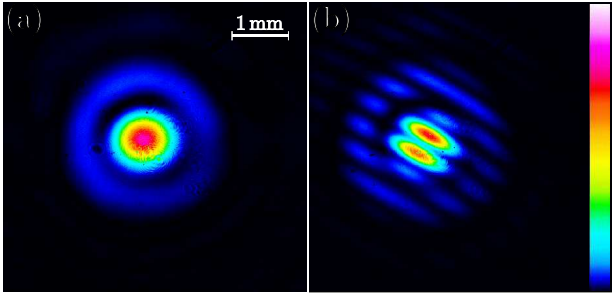}}
\caption[Beam profile]{(a) Beam profile of spontaneous lasing in mercury. (b) The overlapped beam after a Michelson interferometer shows coherence.}
\label{Fig:4}
\end{figure}

Figure \ref{Fig:4}\,(a) shows a measured beam profile of the collimated infrared beam at the exit window of the mercury cell. It should be noted that a second beam propagates in the opposite direction and leaves the cell at the entry window. The input beam parameter products of the fundamental beams are M$^2_{\textrm{uv}}=1.7$ and M$^2_{\textrm{blau}}=1.6$. The resulting dark ring with about 1.4\,mm diameter originates from diffraction at apertures within the vacuum apparatus. To demonstrate the coherence properties of the infrared light a Michelson interferometer is set up. The beam is divided by a beam splitter and overlapped at the same beamsplitter after passing the interferometer arms with arm lengths of about 20\,cm. Figure \ref{Fig:4}\,(b) shows the beam profile after the interferometer with a clearly visible interference pattern. The coherence length estimated from the natural transition linewidth should be about $43$\,cm. 

\begin{figure}[t!]
\centerline{\includegraphics[width=5.8cm,angle=-90]{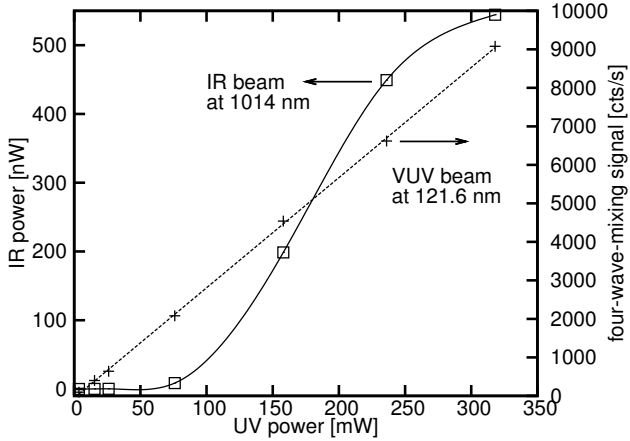}}
\caption[FWMLimit]{Four-wave-mixing power in presence of CTALISE. Infrared powers (boxes) and four-wave-mixing signals (crosses) are shown at different UV powers. The linear fit through the four-wave-mixing data shows that CTALISE is not a limiting process.}
\label{Fig:5}
\end{figure}

Our interest in nonlinear optics in mercury is driven by the need to generate radiation at Lyman-$\alpha$ (121.56\,nm) for future laser-cooling of antihydrogen in a magnetic trap. Mercury vapor is a good candidate for a efficient continuous-wave Lyman-$\alpha$ source by four-wave-mixing \cite{Scheid09}. Amplified spontaneous emission is a possible competing process to Lyman-$\alpha$-generation \cite{Smith88}. To investigate if the four-wave-mixing efficiency is reduced by the amplified spontaneous emission at higher powers we measure the vacuum ultraviolet (VUV) power generated by adding a third laser at 545\,nm wavelength (see Figure \ref{Fig:1}) in the presence of CTALISE. For details of the VUV generation see \cite{Scheid09}. Figure \ref{Fig:5} shows a linear dependence of the Lyman-$\alpha$ power on the UV power. In a four-wave-mixing process the power generated is proportional to the power in each of the fundamental beams \cite{Smith88}. The sub-microwatt of IR lasing does not significantly affect the power in the fundamental beams, and thus the four-wave-mixing is unaffected by the weak IR lasing. The measurement was done at a UV detuning of \mbox{-150\,GHz}, a mercury cell temperature of 153\,$^\circ$C (N$_{\textrm{Hg}}=7.1\times 10^{22}$m$^{-3}$), a buffer gas pressure of 0\,mbar and using the two-photon resonant scheme. The laser powers are 98\,mW in the blue and 189\,mW in the green.

In conclusion, continuous-wave two-photon absorption laser induced stimulated emission (CTALISE) is realized for the first time. The threshold condition of different isotopes is investigated. Efficient pumping to the upper laser level was performed with a double one-photon pumping scheme. 

We gratefully acknowledge support by the BMBF and the DFG.


\end{document}